\documentclass[prl,aps,twocolumn,superscriptaddress]{revtex4}
\usepackage{epsfig}
\usepackage[draft]{hyperref}

\begin{document}
\title{Transport of indirect excitons in ZnO quantum wells}

\author{Y.~Y.~Kuznetsova}
\affiliation{
Department of Physics, University of California at San Diego, La Jolla, CA 92093-0319
}
\author{F.~Fedichkin}
\affiliation{
Laboratoire Charles Coulomb, Universit{\' e} de Montpellier, CNRS, UMR 5221, F-34095 Montpellier, France
}
\author{P.~Andreakou}
\affiliation{
Laboratoire Charles Coulomb, Universit{\' e} de Montpellier, CNRS, UMR 5221, F-34095 Montpellier, France
}
\author{E.~V.~Calman}
\affiliation{
Department of Physics, University of California at San Diego, La Jolla, CA 92093-0319
}
\author{L.~V.~Butov}
\affiliation{
Department of Physics, University of California at San Diego, La Jolla, CA 92093-0319
}
\author{P.~Lefebvre}
\affiliation{
Laboratoire Charles Coulomb, Universit{\' e} de Montpellier, CNRS, UMR 5221, F-34095 Montpellier, France
}
\author{T.~Bretagnon}
\affiliation{
Laboratoire Charles Coulomb, Universit{\' e} de Montpellier, CNRS, UMR 5221, F-34095 Montpellier, France
}
\author{T.~Guillet}
\affiliation{
Laboratoire Charles Coulomb, Universit{\' e} de Montpellier, CNRS, UMR 5221, F-34095 Montpellier, France
}
\author{M.~Vladimirova}
\affiliation{
Laboratoire Charles Coulomb, Universit{\' e} de Montpellier, CNRS, UMR 5221, F-34095 Montpellier, France
}
\author{C.~Morhain}
\affiliation{
Centre de Recherche sur l'H{\'e}t{\'e}ro-Epitaxie et ses Applications, Centre National de la Recherche Scientifique (CRHEA-CNRS), Rue B. Gregory, F-06560 Valbonne Sophia Antipolis, France
}
\author{J.-M.~Chauveau}
\affiliation{
Centre de Recherche sur l'H{\'e}t{\'e}ro-Epitaxie et ses Applications, Centre National de la Recherche Scientifique (CRHEA-CNRS), Rue B. Gregory, F-06560 Valbonne Sophia Antipolis, France
}
\affiliation{
University of Nice Sophia Antipolis, Parc Valrose, F-06102 Nice Cedex 2, France
}

\begin{abstract}
We report on spatially- and time-resolved emission measurements and observation of transport of indirect excitons in ZnO/MgZnO wide single quantum wells.
\end{abstract}
\date{\today}
\maketitle

An indirect exciton (IX) in a semiconductor quantum well (QW) structure is composed of an electron and a hole confined to spatially separated QW layers. IXs were realized in wide single quantum wells (WSQW)~\cite{Miller85, Polland85, Lefebvre04, Morhain05} and in coupled quantum wells (CQW) \cite{Islam87, Alexandrou90, Zrenner92, Butov95} using a variety of QW materials including GaAs~\cite{Miller85, Polland85, Islam87, Alexandrou90}, AlAs~\cite{Zrenner92}, InGaAs~\cite{Butov95}, GaN~\cite{Lefebvre04}, and ZnO~\cite{Morhain05}. Lifetimes of IXs can exceed lifetimes of regular direct excitons by orders of magnitude~\cite{Polland85, Lefebvre04, Morhain05, Alexandrou90, Zrenner92}. Their long lifetimes allow IXs to travel over large distances before recombination, providing the opportunity to study exciton transport by optical imaging~\cite{Fedichkin15, Hagn95, Butov98, Larionov00, Gartner06, Ivanov06, Lazic14} and explore excitonic circuit devices based on exciton transport, see~\cite{Andreakou14} and references therein.

Materials with a high IX binding energy allow extending the operation of the excitonic devices to high temperatures~\cite{Grosso09, Akselrod14, Fogler14}. Furthermore, such materials can allow the realization of high-temperature coherent states of IXs~\cite{Fogler14}. These properties make materials with robust IXs particularly interesting. However, so far, studies of IX transport mainly concerned GaAs-based CQW. In this paper, we probe transport of IXs in ZnO/MgZnO WSQW structures. IXs in these structures are much more robust than in GaAs structures: their binding energy $\sim 30$~meV~\cite{Morhain05} is considerably higher than that in GaAs/AlGaAs and GaAs/AlAs CQW ($\sim 4$ and $\sim 10$~meV, respectively~\cite{Szymanska03, Zrenner92}). The binding energy of IXs is smaller than that of excitons in bulk ZnO ($\sim 60$~meV), however it is large enough to make the IXs stable at room temperature. At the same time, the measurements reported in this work show that transport lengths of IXs in WSQW ZnO structures reach $\sim4$~$\mu$m. In comparison, for excitons in bulk ZnO and direct excitons in ZnO structures, transport lengths are within $\sim0.2$~$\mu$m~\cite{Noltemeyer2012, Friede2015}.

In this work, we study polar and semipolar ZnO/MgZnO QW structures. The samples were grown by molecular beam epitaxy as in Refs.~\cite{Morhain05, Chauveau13}. The semipolar structures consist of an $L_{\rm z} = 5$~nm ZnO QW embedded in 80~nm~/~100~nm Mg$_{0.22}$Zn$_{0.78}$O grown on ($10\overline{1}2$) plane of ZnO substrate. The polar structure consists of an $L_{\rm z} = 7.1$~nm ZnO QW embedded in 200~nm~/~100~nm Mg$_{0.22}$Zn$_{0.78}$O grown on c-plane sapphire substrate following the deposition of $1$~$\mu$m thick ZnO templates. The schematic band diagram of the structure is given in Fig.~1a. The charges on the interfaces between ZnO and MgZnO result in a built-in electric field in the structure, which is stronger for polar samples~\cite{Morhain05, Chauveau13}. The built-in electric field pulls the electron and the hole toward opposite borders of the QW, resulting in the spatial separation required for an IX.

\begin{figure}[h]
\centering
\includegraphics[width=2.3in]{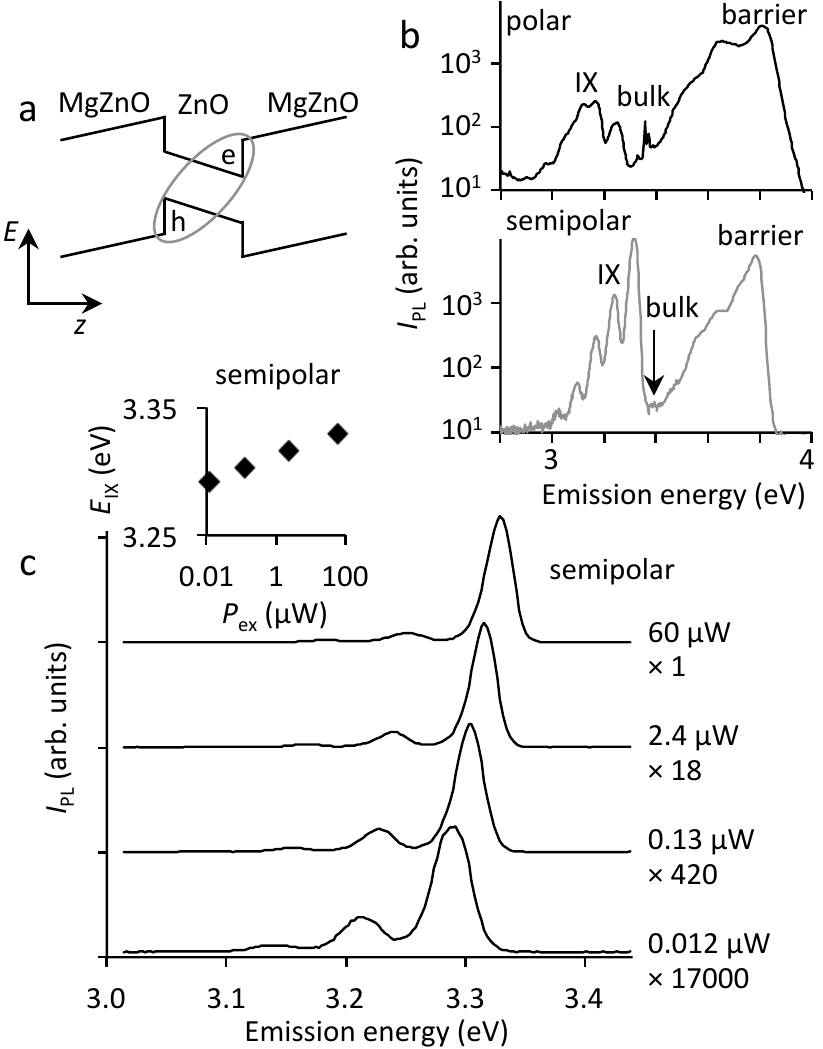}
\label{fig1}
\caption{(a) WSQW ZnO/MgZnO band diagram. The ellipse indicates an IX composed of an electron (e) and a hole (h). (b) Emission spectra at $P_{\rm ex} = 2.4~\mu$W shown on a log scale. IX, ZnO bulk, and MgZnO barrier emission lines are indicated. (c) Emission spectra from the semipolar sample for different $P_{\rm ex}$ shown on a linear scale. Inset: Energy of IX as a function of $P_{\rm ex}$. All spectra are measured at the excitation spot center at $T = 5$~K.}
\end{figure}

In the imaging spectroscopy measurements, excitons were photogenerated by a laser with wavelength 266~nm focused to a 2~$\mu$m excitation spot. %, and the photoluminescence (PL) signal was detected using a nitrogen cooled CCD with spatial resolution $0.8$~$\mu$m. 
In the emission kinetics measurements, excitons were photogenerated by frequency-tripled Ti-Sapphire laser at wavelength 266~nm, pulse duration 150~fs, repetition rate 8~kHz~(0.8~kHz), average power 660~nW~(60~nW), $\sim 100$~$\mu$m excitation spot, and average laser excitation power density $\sim 8$~(0.8)~$\mu$W/cm$^2$ for semipolar (polar) samples. Time-resolved and $x$-energy emission images were measured using a spectrometer equipped with a 150~gr/mm grating blazed at 390~nm. The detector for $x$-energy images was a nitrogen-cooled CCD camera with $1024 \times 256$ pixels and pixel size 26~$\mu$m, which corresponds to 760~nm on the sample surface. Time-resolved measurements of emission kinetics were performed using Hamamatsu streak camera (model C10910 equipped with an S20 photocathode and UV coupling optics), enhanced for ultraviolet detection. This camera can reach an ultimate time resolution of 1~ps, but for the time windows explored in this study, this resolution was 15~ps. The samples were mounted in a cold finger helium cryostat.

Figure~1b presents emission spectra for the structures. The QW confinement results in enhancement of exciton energy. However, for IXs with the separation $d$ between the electron and hole layers, the electric field $F_{\rm z}$ in the sample growth direction reduces the IX energy by $\delta E_{\rm IX} = - edF_{\rm z}$ ($e$ is electron charge). The studied $5$~nm and $7.1$~nm QWs are wide enough that the Stark effect dominates over confinement and the IX energy is below the bulk ZnO energy (Fig.~1b), consistent with earlier measurements~\cite{Morhain05}. The increase of the IX density results in a substantial enhancement of their energy (Fig.~1c). This energy enhancement is specific for IXs and originates from the repulsion between dipole-oriented IXs~\cite{Butov98, Lefebvre04, Morhain05}. The origin of the repulsive behavior is the following: For each pair of IXs separated by the distance $r$ in the QW plane, the distance between their electrons and between their holes ($r$) is smaller than the distance between an electron and a hole ($\sqrt{d^2+r^2}$) and, therefore, the electron-electron and hole-hole repulsion dominates over electron-hole attraction, resulting in repulsion between the IXs. At the lowest IX densities, IX energy lowers to $\sim 0.1$~eV below the bulk ZnO energy. This is comparable to the earlier measured IX energy in ZnO structures~\cite{Morhain05}, indicating a built-in electric field $F_{\rm z} \sim 10^6$~V/cm in the structures. Due to the long lifetimes $\tau$ of IXs, substantial exciton densities $n \propto P_{\rm ex} \tau$ can be achieved using low excitation powers $P_{\rm ex}$ (Fig.~1). Using low $P_{\rm ex}$ minimizes sample heating, which otherwise could lead to a reduction of the emission energy due to band gap reduction or to an energy increase due the thermal energy enhancement. No effect of sample heating was detected in the experiment. An increase of $P_{\rm ex}$ from 0.2 to 60~$\mu$W leads to IX energy enhancement by $\sim 30$~meV for the semipolar sample and $\sim 90$~meV for the polar sample. A stronger energy enhancement for the polar sample is related to longer $\tau$ (Fig.~2), resulting to a higher $n$ for a given $P_{\rm ex}$, and also larger $d$. The lower energy lines in the emission spectra correspond to phonon replicas~\cite{Morhain05}.

\begin{figure}[h]
\centering
\includegraphics[width=2.3in]{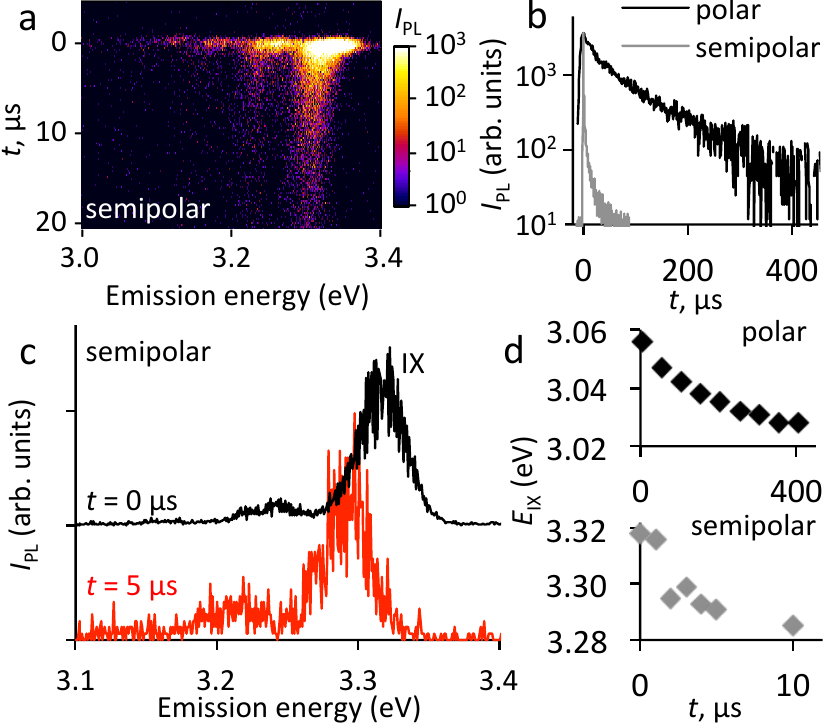}
\label{fig2}
\caption{(a) Emission intensity of IXs in ZnO/MgZnO WSQW as a function of energy and time. (b) Emission intensity integrated over the IX energy as a function of time. (c) Emission spectra of IX at time delays $t=0$ and 5~$\mu$s. (d) Energy of IX as a function of time delay. (a,c) show data from the semipolar sample, while (b,d) show both samples. $T = 8$~K.}
\end{figure}

Figure~2a,b shows the kinetics of IX emission. IXs in the studied structures have long lifetimes. The initial decay time is $\sim 40$~$\mu$s in the polar and $\sim 1$~$\mu$s in semipolar structures (Fig. 2b), with the former value comparable to the earlier measured IX decay time in polar ZnO structures with similar $L_{\rm z}$~\cite{Morhain05}. At long delays the IX decay time reaches~$\sim 100$ and 10~$\mu$s for the polar and semipolar structures, respectively (Fig.~2b). Figure~2c,d shows that the energy of IXs is reduced as a function of time, which originates from the reduction of repulsive IX interaction due to IX density decay~\cite{Lefebvre04, Morhain05}. While the increase of the IX energy with density can be considered as screening of the electric field $F_{\rm z}$ in the structure, the reduction of the IX energy with time delay can be considered as `de-screening' of $F_{\rm z}$. The enhancement of the field $F_{\rm z}$ with time delay $t$ contributes to the observed increase of the IX decay time with $t$~(Fig.~2b).

\begin{figure}[h]
\centering
\includegraphics[width=2.3in]{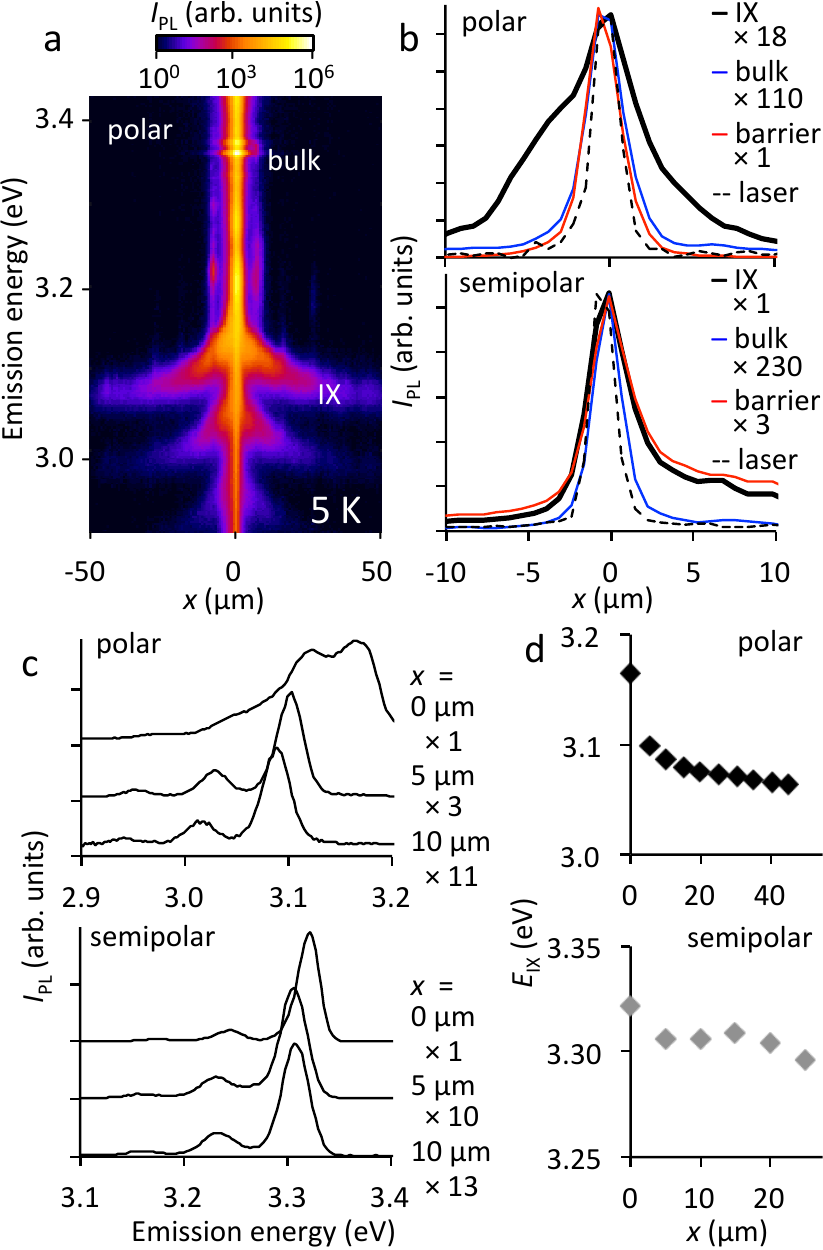}
\label{fig3}
\caption{(a) $x$-energy emission image for the polar structure. (b) Spatial profiles of the intensity of IX emission line in ZnO/MgZnO WSQW (bold black), MgZnO barrier (red), and ZnO bulk (blue). The profile of laser excitation spot is shown by dashed line. (c) Emission spectra at different distances $x$ from the excitation spot center. (d) Energy of IX as a function of $x$. $P_{\rm ex} = 2.4~\mu$W, $T = 5$~K.}
\end{figure}

We measured IX transport by spectrally resolved imaging experiments. An emission image is projected on the spectrometer entrance slit aligned in the $x$-direction through the center of the laser spot. A resulting $x$-energy emission image in the spectrometer exit plane is recorded by a CCD camera. The measured $x$-energy emission images (Fig.~3a) give spectrally resolved spatial profiles of the IX emission. The lower energy emission lines correspond to phonon replicas. The bulk emission is observed higher in energy. Spatial profiles of the IX, bulk, and barrier emission lines shown in Figure~3b are extracted from spectra such as the one pictured in Fig.~3a. Vertical cross-sections of the IX $x$-energy emission image give IX emission spectra at different distances $x$ from the origin (Fig.~3c). Figure~3b and 3d show the intensity and energy, respectively, of the IX emission line vs $x$.

Figures~3a,c,d show that the energy of IXs is reduced with increasing distance $x$ from the excitation spot center. The observed IX energy reductions with $x$ (Fig.~3a,c,d), with the delay time $t$ (Fig.~2c,d), and with the decreasing excitation power $P_{\rm ex}$ (Fig.~1c) have similar origins. These IX energy reductions originate from reduction of the repulsive IX interaction due to lowering IX density with increasing $x$, increasing $t$, and decreasing $P_{\rm ex}$, respectively.

The transport lengths of excitons in bulk ZnO are typically small due to short lifetimes of excitons in the bulk \cite{Noltemeyer2012}. Therefore, spatial profile of the bulk emission is close to the laser excitation profile (Fig.~3b). In contrast, the IX cloud expansion in the polar structure $\sim 3-5$~$\mu$m significantly exceeds the size of the laser excitation spot as well as the barrier and bulk emission expansions, see Fig.~3a,b. The IX cloud expansion, seen from the spatial profile of the IX emission, reveals transport of IXs $\sim 4$~$\mu$m away from the excitation spot. The large transport distances for IXs are achieved due to their long lifetimes (Fig.~2), which are orders of magnitude longer than lifetimes of excitons in bulk ZnO.

\begin{figure}[h]
\centering
\includegraphics[width=2.3in]{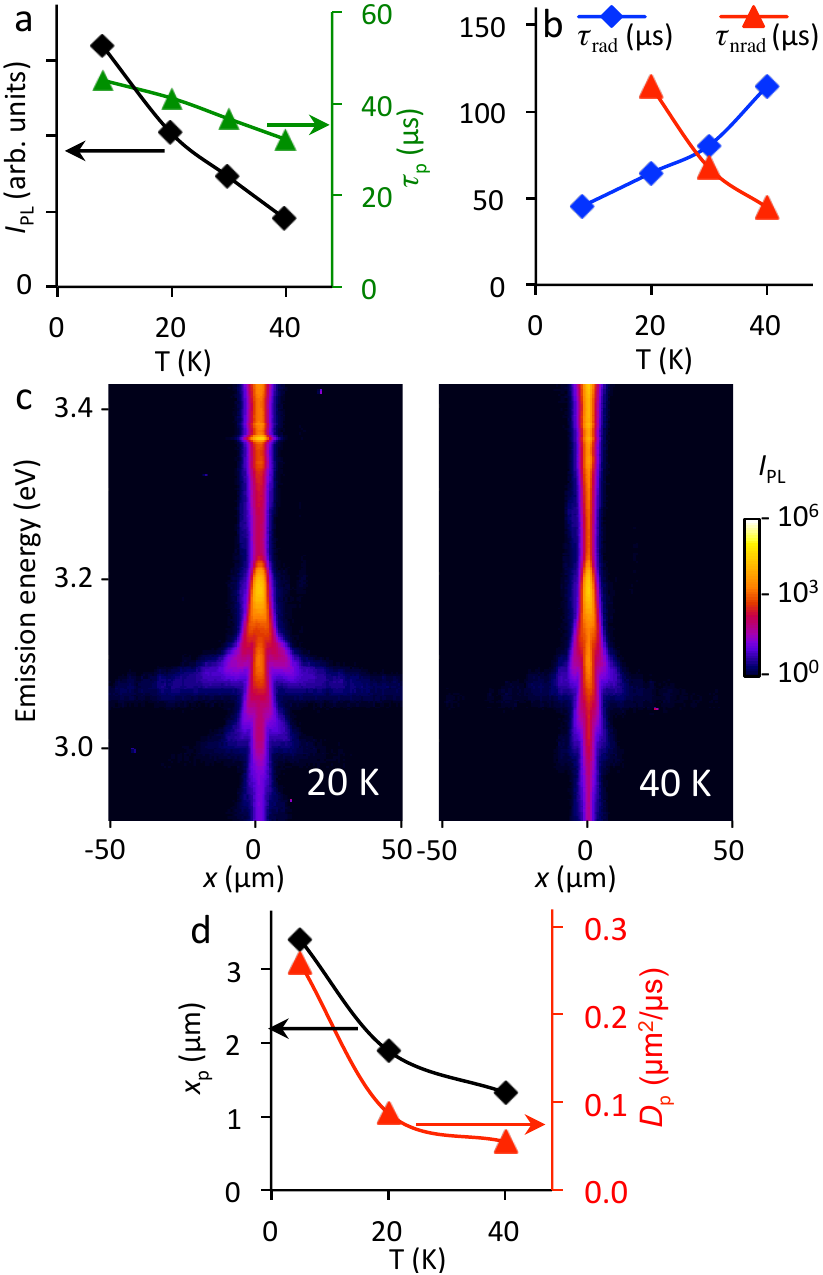}
\label{fig4}
\caption{Temperature dependence in the polar structre of (a) measured spatially and spectrally integrated emission intensity $I$ (black diamonds) and initial decay time $\tau_{\rm p}$ (green triangles), (b) derived radiative $\tau_{\rm rad}$ (blue diamonds) and nonradiative $\tau_{\rm nrad}$ (red triangles) decay times, (c) measured $x$-energy emission image for IXs, and (d) measured IX transport length $x_{\rm p}$ (black diamonds) and estimated IX diffusion coefficient $D_{\rm p} = x_{\rm p}^2/\tau_{\rm p}$ (red triangles). $P_{\rm ex} = 2.4~\mu$W.}
\end{figure}

A rough estimate of the IX diffusion coefficient in the polar structure $D_{\rm p}$ can be made using the measured IX cloud expansion beyond the optical excitation spot $x_{\rm p} \sim 3.5$~$\mu$m (Fig.~3b) and initial exciton decay time $\tau_{\rm p} \sim 40$~$\mu$s~(Fig.~2b) (in this estimate, the initial exciton decay time is used to describe exciton transport in the area where the exciton density drops by half from its initial value): $D_{\rm p} \sim x_{\rm p}^2/\tau_{\rm p} \sim 3 \times 10^{-3}$~cm$^2$/s. It is 3 orders of magnitude smaller than the exciton diffusion coefficient reported in GaAs structures \cite{Ivanov06}. This is likely related to a stronger in-plane disorder potential in the ZnO structures. Note that even in the GaAs structures, the diffusion of IX is quenched when the exciton density is small and screening of the in-plane structure disorder by exciton-exciton interaction is suppressed; this regime is realized at small excitation powers and also near the borders of the exciton cloud \cite{Ivanov06}.

We also performed spectrally resolved imaging experiments for semipolar structure. The results of these measurements are presented in Fig.~3b-d. Both the barrier and IX emission in the semipolar sample are asymmetric. These asymmetries may originate from asymmetries in disorder in the structure and excitation spot shape. A smaller emission asymmetry for polar sample (Fig.~3b) is likely related to a more symmetric excitation spot profile in the experiments with the polar samples. These asymmetries are insignificant for the discussions in the paper. A rough estimate for the semipolar structure using the measured IX cloud expansion beyond the optical excitation spot $x_{\rm sp} \sim 2$~$\mu$m (Fig.~3b) and initial exciton decay time $\tau_{\rm sp} \sim 1$~$\mu$s~(Fig.~2b) gives $D_{\rm sp} \sim x_{\rm sp}^2/\tau_{\rm sp} \sim 4 \times 10^{-2}$~cm$^2$/s, that appears an order of magnitude larger than $D_{\rm p}$. Note however that for the semipolar sample the barrier emission spreads beyond the excitation spot over a distance similar to the IX emission. Therefore, the optically excited carriers expanding in the MgZnO barrier may in principle relax to the lower-energy QW thus feeding QW away from the excitation spot and contributing to the observed IX cloud expansion. Taking this into account should reduce the estimated value of the IX diffusion coefficient for the semipolar sample. Therefore, for the semipolar sample, the measurements provide only the upper boundary for the IX diffusion coefficient $D_{\rm sp} \lesssim 4 \times 10^{-2}$~cm$^2$/s.

We probed the temperature dependence of IX kinetics and transport in the polar structure. Figure~4a shows that both $\tau$ and the integrated emission intensity $I$ of IXs are reduced with temperature. For an estimate of radiative and nonradiative decay times, $\tau_{\rm rad}$ and $\tau_{\rm nrad}$, we use the relations $\tau^{-1} = \tau_{\rm rad}^{-1} + \tau_{\rm nrad}^{-1}$ and $I = G \tau / \tau_{\rm rad}$, where $G$ is the generation rate, and assume that at the lowest studied temperature $\tau_{\rm rad} \ll \tau_{\rm nrad}$ and, as a result, $G \approx I_{\rm 8 K}$. Figure~4b shows that $\tau_{\rm rad}$ increases with temperature, qualitatively consistent with the reduction of occupation of low-energy optically active exciton states \cite{Feldman87}. Figure~4b also shows that $\tau_{\rm nrad}$ is reduced with temperature, indicating that nonradiative processes become stronger. The enhancement of nonradiative recombination rate at high temperatures results in the reduction of exciton decay time and emission intensity (Fig.~4a). Figures~4c,d show that IX transport lengths $x_{\rm p}$ decrease with increasing temperature. This indicates that the development of high-temperature excitonic devices in ZnO structures should be facilitated by the improvement of the structure quality, in particular, suppression of nonradiative recombination and disorder in the structures. The IX diffusion coefficient $D_{\rm p} \sim x_{\rm p}^2/\tau_{\rm p}$ decreases with temperature (Fig.~4d).

In conclusion, we performed imaging spectroscopy measurements of IX in ZnO/MgZnO wide single quantum well structures, observed transport of IX, and estimated the IX diffusion coefficient.

We thank Frances Hellman for valuable discussions and contributions at the earlier stage of studies of IX in ZnO structures. This work was supported by NSF and by EU ITN INDEX. Y.Y.K. was supported by an Intel fellowship.

\end{document}